\documentstyle[amssymb]{article}

\input xy
\xyoption{all}

\CompileMatrices

\begin{document}

\title{Nonlocal lagrangians and mass generation for gauge fields}

\author{A. Sevostyanov \\
Department of Mathematical Sciences \\
University of Aberdeen  \\ Aberdeen AB24 3UE, United Kingdom \\
e-mail: seva@maths.abdn.ac.uk \\ phone: +44-1224-272755, fax:
+44-1224-272607}

\maketitle

\begin{abstract}
In this paper we study the nonabelian, gauge invariant and
asymptotically free quantum gauge theory with a mass parameter
introduced in \cite{S}. We develop the Feynman diagram technique,
calculate the mass and coupling constant renormalizations and the
effective action at the one--loop order. Using the BRST technique
we also prove that the theory is renormalizable within the
dimensional regularization framework.
\end{abstract}



\section{Introduction}
In paper \cite{S} we introduced a new classical gauge invariant
nonlocal lagrangian generating a local quantum field theory which
is reduced to several copies of the massive vector field when the
coupling constant vanishes. This unexpected phenomenon is similar
to that for the Faddeev-Popov determinant in case of the quantized
Yang-Mills field. Recall that being a priori nonlocal quantity the
Faddeev-Popov determinant can be made local by introducing
additional anticommuting ghost fields and applying a formula for
Gaussian integrals over Grassmann variables. Similarly, in case of
the lagrangian suggested in \cite{S} one can introduce extra ghost
fields and make the expression for the generating function of the
Green functions local using tricks with Gaussian integrals.

In this paper we develop the Feynman diagram technique, calculate
the mass and coupling constant renormalizations and the effective
action at the one--loop order for the theory introduced in
\cite{S}. Using the BRST technique we also prove that the theory
is renormalizable within the dimensional regularization framework.

\section{Preliminaries}
First we fix the notation used throughout this paper. Let $G$ be a
compact simple Lie group, $\mathfrak g$ its Lie algebra with the
commutator denoted by $[\cdot, \cdot ]$. Denote by $\textrm{tr}$
the Killing form on $\mathfrak g$ with the opposite sign. Recall
that $\textrm{tr}$ is a nondegenerate invariant under the adjoint
action scalar product on $\mathfrak g$. Let $t^a$, $a=1,\ldots
,\textrm{dim}{\mathfrak g}$ be a linear basis of $\mathfrak g$
normalized in such a way that $\textrm{tr}(t^a,t^b)=\delta^{ab}$.
Note that for technical reasons the normalizations of scalar
products in this paper are slightly different from \cite{S} and
from the standard ones (see \cite{IZ}).

Since the tensor formulas that will appear in this paper are even
more complicated than in case of the Yang-Mills field we shall
frequently use the invariant tensor notation and a shorten
notation for scalar products of tensors. Denote by
$\Omega^k({\mathfrak g})$ the space of $\mathfrak g$--valued
differential k-forms on the standard Minkowski space equipped with
the metric $g_{\mu\nu}$, $g_{00}=1$, $g_{ii}=-1$ for $i=1,2,3$,
and $g_{ij}=0$ for $i\neq j$.

We shall use the scalar product $<\cdot, \cdot>$ on the space
$\Omega^k({\mathfrak g})$ defined by
\begin{equation}\label{prod}
<\omega, \omega'>=\int \textrm{tr}(\omega\wedge,
*\omega'),~\omega, \omega'\in \Omega^k({\mathfrak g}),
\end{equation}
where $*$ stands for the Hodge star operation associated to the
metric on  the Minkowski space, and we evaluate the Killing form
on the values of $\omega_1$ and $*\omega_2$ and also take their
exterior product.

Let $A\in \Omega^1({\mathfrak g})$ be the $\mathfrak g$-valued
gauge field (connection one-form on the Minkowski space),
$$
A=A_\mu dx^\mu=A_\mu^at^a dx^\mu,
$$
and $F\in \Omega^2({\mathfrak g})$ the strength tensor (curvature)
of $A$,
$$
F=dA-\frac{\textrm{g}}{2}[A\wedge,A].
$$
Here $\textrm{g}$ is a coupling constant and as usual we denote by
$[A\wedge, A]$ the operation which takes the exterior product of
$\mathfrak g$-valued 1-forms and the commutator of their values in
$\mathfrak g$. In terms of components we have
$$
F=F_{\mu\nu}dx^\mu \wedge dx^\nu=\frac{1}{2}(\partial_\mu A_\nu
-\partial_\nu A_\mu -\textrm{g}[A_\mu,A_\nu])dx^\mu \wedge dx^\nu.
$$

We shall also need the covariant derivative operator
$d_A:\Omega^k({\mathfrak g})\rightarrow \Omega^{k+1}({\mathfrak
g})$ associated to the connection $A$,
$$
d_A\omega =d\omega-\textrm{g}[A\wedge, \omega].
$$
The operator $d_A^*$ adjoint to $d_A$ with respect to scalar
product (\ref{prod}) is equal to $(-1)^{k(4-k)+1}*d_A*$.

The gauge group of $G$-valued functions $g(x)$ defined on the
Minkowski space acts on the gauge field $A$ by
\begin{equation}\label{gaugen}
A \mapsto \frac{1}{\textrm{g}}dg g^{-1}+gAg^{-1}.
\end{equation}
This action generates transformation laws for the covariant
derivative and the strength tensor,
\begin{eqnarray}
d_A \mapsto gd_A g^{-1}, \label{trcov}\\
F \mapsto gFg^{-1}. \label{trf}
\end{eqnarray}
In formula (\ref{trcov}) we assume that the gauge group acts on
tensor fields according to the representation of the group $G$
induced by that of the Lie algebra $\mathfrak g$, and the r.h.s.
of (\ref{trcov}) should be regarded as the composition of
operators.

We shall also need a covariant rough d'Alambert operator
$\square_A$ associated to the gauge field $A$,
$$
\square_A=D_\mu D^\mu,~D_\mu=\partial_\mu-\textrm{g}A_\mu,
$$
Here and thereafter we use the standard convention about
summations and lowering tensor indexes with the help of the
standard metric on the Minkowski space.

The covariant d'Alambert operator can be applied to any tensor
field defined on the Minkowski space and taking values in a
representation space of the Lie algebra $\mathfrak g$, the
$\mathfrak g$-valued gauge field $A$ acts on the tensor field
according to that representation. Note that the operator
$\square_A$ is scalar, i.e. it does not change types of tensors.

Formula (\ref{trcov}) implies that the covariant d'Alambert
operator is transformed under gauge action (\ref{gaugen}) as
follows
\begin{equation}\label{trdal}
\square_A \mapsto g\square_A g^{-1}.
\end{equation}

For the zero connection $A=0$ we simply write $d_0=d$ and
$\square_0=\square$, the usual d'Alambert operator.

Using the tensor notation the ``localized'' action for the massive
gauge field $A$ introduced in \cite{S} can defined by the formula
\begin{eqnarray}\label{act1}
S'= -\frac{1}{2}<F,F>+\frac{1}{32}<\square_A\Phi,\Phi>
-\frac{m}{4}<\Phi,F>+
\\*+\sum_{i=1}^3<d_A\overline{\eta}_i, d_A\eta_i>. \nonumber
\end{eqnarray}
Here $\Phi \in \Omega^{2}({\mathfrak g})$ is a bosonic ghost field
in the adjoint representation of $\mathfrak g$; $\eta_i,
\overline{\eta}_i$, $i=1,2,3$ are pairs of anticommuting scalar
ghost fields in the adjoint representation of $\mathfrak g$. In
formula (\ref{act1}) $\textrm{g}$ should be regarded as a coupling
constant and $m$ is a mass parameter. From (\ref{trcov}),
(\ref{trf}) and (\ref{trdal}) it follows that action (\ref{act1})
is invariant under gauge transformations (\ref{gaugen}).

In formula (\ref{act1}) we use normalizations slightly different
from those introduced in \cite{S}. For instance, the coefficient
in front of the last term in (\ref{act1}) is different from that
which we used in \cite{S}. But this coefficient is not important
for the study of renormalizability of the theory.

To define the Green functions corresponding to the gauge invariant
action $S'$ we have to add to action (\ref{act1}) a gauge fixing
term $S_{gf}$. As it was observed in \cite{S} the most convinient
choice of the gauge fixing term is
$$
S_{gf}=-\frac{\lambda}{2}<d^*A,d^*A>-\frac{m^2}{2}<\square^{-1}d^*A,d^*A>+
<d\overline{\eta},d_A\eta>,
$$
where $\eta, \overline{\eta}$ is a pair of anticommuting scalar
ghost fields in the adjoint representation of $\mathfrak g$, and
we introduced dimensionless parameter $\lambda$ to consider
different gauge fixing conditions (in \cite{S} we only discussed
the case when $\lambda=1$). In this paper we shall also assume
that all inverse d'Alambert operators are of Feynman type.

The total action $S=S'+S_{gf}$ that should be used in the
definition of the Green functions takes the form
\begin{eqnarray}
S= -\frac{1}{2}<F,F>+\frac{1}{32}<\square_A\Phi,\Phi>
-\frac{m}{4}<\Phi,F>+<d\overline{\eta},d_A\eta>+ \label{actloc}
\\*+\sum_{i=1}^3<d_A\overline{\eta}_i, d_A\eta_i>
-\frac{\lambda}{2}<d^*A,d^*A>-\frac{m^2}{2}<\square^{-1}d^*A,d^*A>.
\nonumber
\end{eqnarray}

Now consider the expression for the generating function $Z(J)$ of
the Green functions for the theory with action (\ref{actloc}) via
a Feynman path integral,
\begin{equation}\label{gener}
Z(J)=\int{\mathcal D}(A){\mathcal D}(\Phi){\mathcal
D}(\eta){\mathcal D}(\overline{\eta})\prod_{i=1}^3{\mathcal
D}(\eta_i){\mathcal D}(\overline{\eta}_i) \exp \{i(S+<J,A>)\},
\end{equation}
where $J$ is the source for $A$. Here and thereafter we assume
that the measure in Feynman path integrals is suitably normalized.
We also assume that the Feynman path integral over $\Phi$ in
(\ref{gener}) is only taken with respect to the linearly
independent components $\Phi_{\mu\nu}$, $\mu < \nu$ of the
skew-symmetric tensor $\Phi_{\mu\nu}$.

Observe that in the r.h.s. of formula (\ref{gener}) all the
integrals over the ghost fields are Gaussian. The Gaussian
integrals can be explicitly evaluated (see \cite{S} for details).
This yields
\begin{eqnarray}\label{act}
Z(J)=\int{\mathcal D}(A)\det (d^*d_A|_{\Omega^0(\mathfrak g)})\exp
\{ i (-\frac{1}{2}<F,F>-\frac{m^2}{2}<\square^{-1}_AF,F>+ \\
+<J,A>-\frac{\lambda}{2}<d^*A,d^*A>-\frac{m^2}{2}<\square^{-1}d^*A,d^*A>)
\}.\nonumber
\end{eqnarray}

The r.h.s. of (\ref{act}) looks like the generating function of
the Green functions for the Yang-Mills theory with an extra
nonlocal term, $-\frac{m^2}{2}<\square^{-1}_AF,F>$, and in a
generalized gauge (see \cite{FS}, Ch 3, \S 3). Although  the
action $S_m$ that appears in formula (\ref{act}),
\begin{eqnarray}\label{anonloc}
S_m=-\frac{1}{2}<F,F>-\frac{m^2}{2}<\square^{-1}_AF,F>
-\frac{\lambda}{2}<d^*A,d^*A>-
\\
\qquad \qquad \qquad \qquad \qquad \qquad \qquad \qquad
-\frac{m^2}{2}<\square^{-1}d^*A,d^*A>, \nonumber
\end{eqnarray}
is not local the generating function $Z(J)$ for this action is
equal to that for action (\ref{actloc}). It is the phenomenon that
was observed in \cite{S}.

The purpose of this paper is to study generating function
(\ref{gener}) in detail. In the next section we shall discuss the
Feynman diagram technique for action (\ref{actloc}).

\section{Feynman diagram technique}

In order to develop perturbation theory for generating function
(\ref{gener}) we have to introduce additional sources $I\in
\Omega^2(\mathfrak g)$, $\xi$, $\overline\xi$, $\xi^i$,
$\overline{\xi}^i$, $i=1,2,3$ for the ghost fields $\Phi$, $\eta$,
$\overline\eta$, $\eta_i$, $\overline{\eta}_i$, $i=1,2,3$,
respectively. We also need the full generating function
$Z(J,I,\xi, \overline\xi,\xi^i,\overline{\xi}^i)$ for the Green
functions,
\begin{eqnarray}
Z(J,I,\xi, \overline\xi,\xi^i,\overline{\xi}^i)=\int{\mathcal
D}(A){\mathcal D}(\Phi){\mathcal D}(\eta){\mathcal
D}(\overline{\eta})\prod_{i=1}^3{\mathcal D}(\eta_i){\mathcal
D}(\overline{\eta}_i)\times \nonumber \\
\times \exp
\{i(S+<J,A>+\frac{1}{4}<I,\Phi>+<\overline\xi,\eta>+<\xi,
\overline\eta>+ \label{zf} \\
+\sum_{i=1}^3(<\overline{\xi}^i,\eta_i>+<\xi^i,
\overline{\eta}_i>))\}. \nonumber
\end{eqnarray}
Recall that the connected Green functions are defined as the
coefficients of the generating series $G(J,I,\xi,
\overline\xi,\xi^i,\overline{\xi}^i)$ related to $Z(J,I,\xi,
\overline\xi,\xi^i,\overline{\xi}^i)$ by
$$
Z(J,I,\xi, \overline\xi,\xi^i,\overline{\xi}^i)=\exp (G(J,I,\xi,
\overline\xi,\xi^i,\overline{\xi}^i)).
$$

We shall first calculate the free propagators for the theory with
generating function (\ref{gener}). The free propagators are the
connected Green functions corresponding to the free generating
function $Z_0(J,I,\xi, \overline\xi,\xi_i,\overline{\xi}_i)$,
\begin{eqnarray}
Z_0(J,I,\xi, \overline\xi,\xi^i,\overline{\xi}^i)=\int{\mathcal
D}(A){\mathcal D}(\Phi){\mathcal D}(\eta){\mathcal
D}(\overline{\eta})\prod_{i=1}^3{\mathcal D}(\eta_i){\mathcal
D}(\overline{\eta}_i)\times \nonumber \\
\times \exp
\{i(S_0+<J,A>+\frac{1}{4}<I,\Phi>+<\overline\xi,\eta>+<\xi,
\overline\eta>+ \label{zo} \\
+\sum_{i=1}^3(<\overline{\xi}^i,\eta_i>+<\xi^i,
\overline{\eta}_i>))\}, \nonumber
\end{eqnarray}
where
\begin{eqnarray}
S_0= -\frac{1}{2}<dA,dA>+\frac{1}{32}<\square\Phi,\Phi>
-\frac{m}{4}<\Phi,dA>+<d\overline{\eta},d\eta>+ \label{actloco}
\\*+\sum_{i=1}^3<d\overline{\eta}_i, d\eta_i>
-\frac{\lambda}{2}<d^*A,d^*A>-\frac{m^2}{2}<\square^{-1}d^*A,d^*A>
\nonumber
\end{eqnarray}
is the unperturbed action corresponding to (\ref{actloc}).

Feynman path integral (\ref{zo}) is Gaussian, and we can
explicitly evaluate it. First we integrate over the ghost fields
in (\ref{zo}). This yields
\begin{eqnarray}
Z_0(J,I,\xi, \overline\xi,\xi^i,\overline{\xi}^i)=\int{\mathcal
D}(A)\exp
\{i(-\frac{1}{2}<(d^*d+dd^*) A,A>- \qquad \qquad \qquad \nonumber \\
\label{zo1}
-\frac{(\lambda-1)}{2}<dd^*A,A>-\frac{m^2}{2}<(d^*d+dd^*)\square^{-1}
A,A>-<m\square^{-1}d A,I>+ \\
+<J,A>-\frac{1}{2}<\square^{-1}I,I>+
<\square^{-1}\overline\xi,\xi>+\sum_{i=1}^3
<\square^{-1}\overline{\xi}^i,\xi^i>)\}. \nonumber
\end{eqnarray}

Now using the Hodge formula, $dd^*+d^*d=-\square$, in the exponent
in (\ref{zo1}) and integrating over $A$ we finally obtain
\begin{eqnarray}
Z_0(J,I,\xi, \overline\xi,\xi^i,\overline\xi^i)=\exp
\{i(-\frac{1}{2}<(\square+m^2+(1-\lambda)dd^*)^{-1}J,J>- \nonumber \\
-\frac{1}{2}<(\square^{-1}+m^2\square^{-2}dd^*(\square+m^2+(1-\lambda)dd^*)^{-1})I,I>-\label{prop}
\\
-<m(\square+m^2+(1-\lambda)dd^*)^{-1}\square^{-1}d^*I,J>+
\nonumber
\\
+<\square^{-1}\overline\xi,\xi>+\sum_{i=1}^3
<\square^{-1}\overline{\xi}^i,\xi^i>)\}, \nonumber
\end{eqnarray}
where all the inverse d'Alambert operators are of Feynman type. We
see that the free propagator for the gauge field $A$ coincides
with that for a massive field. When $\lambda=1$ it is reduced to
the propagator for the free massive vector field,
$i(\square+m^2)^{-1}$. The fact crucial for the mass generation is
that free action (\ref{actloco}) contains a cross term,
$-\frac{m}{4}<\Phi,dA>$, which generates mass in the free
propagator of the gauge field. Note that the free propagator is
also not diagonal. It contains the cross term $
<m(\square+m^2+(1-\lambda)dd^*)^{-1}\square^{-1}d^*I,J>$.

We shall develop perturbation theory in the momentum space for the
Fourier transforms of the Green functions. We normalize the
Fourier transform as follows
$$
\widehat f(k)=\frac{1}{(2\pi)^4}\int e^{-ik\cdot x} f(x)d^4x.
$$
The Fourier transforms of the propagators arising from
(\ref{prop}), along with the corresponding diagrammatic symbols,
are summarized in the following table:

\begin{eqnarray*}
 \xymatrix@=1.5cm{  J^{\mu}_a {~}\ar@{{*}~}[r]^<*+{\mu}_<*+{a}& *=0{>} \ar@{~{*}}[r]^>*+{\nu}_>*+{b}^<*+{k}&
{~}J^{\nu}_b} \qquad  i(\square+m^2+(1-\lambda)dd^*)^{-1}= \qquad \qquad \qquad \qquad \\
 \qquad \qquad \qquad  \qquad \quad  =i\delta^{ab}\left(
-\frac{g_{\mu\nu}}{k^2-m^2+i\varepsilon}+
\frac{(1-\lambda^{-1})k_\mu
k_\nu}{(k^2-m^2+i\varepsilon)(k^2-\frac{m^2}{\lambda}+i\varepsilon)}\right)
\end{eqnarray*}
\begin{eqnarray*}
 \hspace{0.5cm}\xymatrix@1@=1.5cm{ I^{\mu \gamma}_a {~} \ar@2{{*}-}[r]^<<*+{\mu < \gamma}_<*+{a}& *=0{>} \ar@2{-{*}}[r]^>>*+
 {\nu < \delta}_>*+{b}^<*+{k} & {~~} I^{\nu \delta}_b}
 \quad i(\square^{-1}+m^2\square^{-2}dd^*(\square+m^2+(1-\lambda)dd^*)^{-1})=\qquad
 \qquad
 \\
 =\frac{-4i\delta^{ab}}{k^2+i\varepsilon} \left[ g_{\mu \nu}g_{\gamma
 \delta}+\frac{m^2}{(k^2+i\varepsilon)(k^2-m^2+i\varepsilon)}
 \left(g_{\mu \nu}k_\gamma k_\delta -g_{\gamma \nu}k_\mu
 k_\delta -g_{\mu \delta} k_\gamma k_\nu+g_{\gamma \delta}k_\mu
 k_\nu\right)\right]
\end{eqnarray*}

\begin{eqnarray}
\hspace{-1.5cm}\xymatrix@1@=1.5cm{ J^{\mu}_a {~}
\ar@{{*}~}[r]^<*+{\mu}_<*+{a}|*=0{>} & *=0{\bullet}
\ar@2{-{*}}[r]^>>*+{\nu < \gamma}_>*+{b}^<*+{k}|*=0{>}& {~~}I^{\nu
\gamma}_b } \qquad
i m(\square+m^2+(1-\lambda)dd^*)^{-1}\square^{-1}d^* = \label{cross1} \\
=-\frac{2m
\delta^{ab}}{(k^2+i\varepsilon)(k^2-m^2+i\varepsilon)}\left(
g_{\mu \nu}k_\gamma -g_{\mu \gamma} k_\nu \right) \nonumber
\end{eqnarray}

\begin{eqnarray*}
\hspace{-1.3cm}\xymatrix@=1.5cm{ \xi_a \ar@{{*}--}[r]_<*+{a}&
*=0{>} \ar@{--{*}}[r]_>*+{b}^<*+{k}&{~} \xi_b }\qquad
-i\square^{-1}= \frac{i\delta^{ab}}{k^2+i\varepsilon} \qquad
\qquad \qquad \qquad \quad
\end{eqnarray*}

\begin{eqnarray*}
\hspace{-1.3cm}\xymatrix@=1.5cm{ \xi_a^i \ar@{{*}--}[r]_<*+{a}&
*=0{>} \ar@{--{*}}[r]_>*+{b}^<*+{k}_<<{i}&{~} \xi_b^i }\qquad
-i\square^{-1}= \frac{i\delta^{ab}}{k^2+i\varepsilon} \qquad
\qquad \qquad \qquad \quad
\end{eqnarray*}
In the diagrams above we indicate the sources coupled to the
corresponding propagators in formula (\ref{prop}). We assume that
in coordinate formulas containing summations over indexes of
skew-symmetric (2,0)-type tensors all summations are only taken
over the indexes corresponding to the independent components of
the skew-symmetric tensors; for instance, in formulas containing
summations over the indexes of the source $I^{\mu \nu}_a$ we take
sums over $\mu <\nu$. This gives some extra numerical factors in
the formulas for the propagators. For instance, we have
$$
<\omega,\omega'>=
4\int\sum_{\mu<\nu}\textrm{tr}(\omega_{\mu\nu},\omega'^{\mu\nu})d^4x,~\omega,\omega'\in
\Omega^2(\mathfrak g).
$$
In the formulas for the Fourier
transforms of the propagators given above we also omit, as usual,
the delta functions of the total external momenta.

The perturbation theory for generating series (\ref{gener}) is
based on the following formula
\begin{eqnarray}\label{perturbat}
Z(J)= \qquad \qquad \qquad \qquad \qquad \qquad \qquad \qquad
\qquad \qquad \qquad \qquad \qquad \qquad  \qquad \qquad
\\
=\exp\left\{iV\left( \frac{\delta}{i\delta
J},\frac{\delta}{i\delta I},\frac{\delta}{i\delta
\xi},\frac{\delta}{i\delta \overline\xi},\frac{\delta}{i\delta
\xi^i},\frac{\delta}{i\delta \overline{\xi}^i}\right)
\right\}\left|_{I=\xi=\overline\xi=\xi^i=\overline{\xi}^i=0}Z_0(J,I,\xi,
\overline\xi,\xi^i,\overline{\xi}^i), \right. \nonumber
\end{eqnarray}
where
$$
V(A,\Phi,\eta,\overline{\eta},\eta_i,\overline{\eta}_i)=
S(A,\Phi,\eta,\overline{\eta},\eta_i,\overline{\eta}_i)-
S_0(A,\Phi,\eta,\overline{\eta},\eta_i,\overline{\eta}_i)
$$
is the perturbation of the free action $S_0$, and in formula
(\ref{perturbat}) all variational derivatives with respect to the
Grassmann variables $\overline\xi$, $\overline\xi_i$ ($\xi$,
$\xi_i$) are right (left), respectively.

To complete our study of the perturbation theory based on formula
(\ref{perturbat}) we have to calculate the contributions of the
vertices corresponding to the terms of the perturbation $iV$.
These terms and their contributions evaluated in the momentum
representation, along with the corresponding diagrammatic symbols,
are listed below.

\vskip -2cm
$$  \xymatrix@!@=1.5cm{ & \ar@{<~}[d]_<*+{\mu}^<*+{a}^*+{p}& \\
&  *=0{\bullet}  & \\
\ar@{<~}[ur]^<*+{\nu}_<*+{b}^*+{q} & & \ar@{<~}[ul]_<*+{\rho}^<*+{c}_*+{r}}  \begin{array}{l} \\
\\
\\
\\
\\
\frac{i\textrm{g}}{2}<dA,[A\wedge,A]>,  \\
  \\
  \\
  \textrm{g}C^{abc}(2\pi)^4\delta(p+q+r)( g_{\mu
\nu}(p-q)_\rho+g_{\nu \rho}(q-r)_\mu +  \\
 \\
\qquad \qquad \qquad \qquad \qquad \qquad \qquad \qquad \quad
+g_{\rho \mu}(r-p)_\nu)
\end{array}
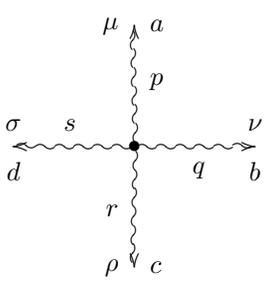
$$
\vskip -2cm
$$
\hspace{1cm}\xymatrix@!@=1.5cm{ & \ar@{<~}[d]_<*+{\mu}^<*+{a}^*+{p}& \\
 \ar@{<~}[r]^<*+{\sigma}_<*+{d}^*+{s} &  *=0{\bullet}  & \ar@{<~}[l]_<*+{\nu}^<*+{b}^*+{q}\\
 & \ar@{<~}[u]^<*+{\rho}_<*+{c}^*+{r} & } \begin{array}{l} \\
\\
\\
\\
\\
-\frac{i\textrm{g}^2}{8}<[A\wedge,A],[A\wedge,A]>, \\
 \\
 \\
 -i\textrm{g}^2(2\pi)^4\delta(p+q+r+s)(
 C^{eab}C^{ecd}(g_{\mu \rho}g_{\nu \sigma}-g_{\mu \sigma}g_{\nu
 \rho})+ \\
 \\
 +C^{eac}C^{edb}(g_{\mu \sigma}g_{\rho \nu}-g_{\mu \nu}g_{\rho
 \sigma})+C^{ead}C^{ebc}(g_{\mu \nu}g_{\sigma \rho}-g_{\mu \rho}g_{\sigma \nu}))\end{array}
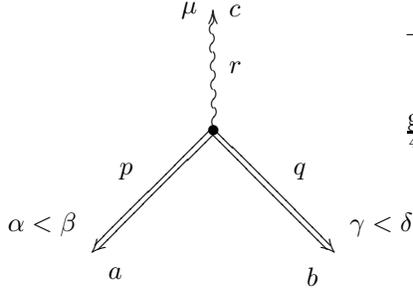
$$
\vskip -2cm
$$
\xymatrix@!@=1.5cm{ & \ar@{<~}[d]^*+{r}_<*+{\mu}^<*+{c}& \\
&  *=0{\bullet}  & \\
\ar@2{<-}[ur]^*+{p}^<*+{\alpha < \beta}_<*+{a} & &
\ar@2{<-}[ul]_*+{q}_<*+{\gamma < \delta}^<*+{b}}
\hspace{-0.5cm}\begin{array}{l} \\
\\
\\
\\
\\
-\frac{i\textrm{g}}{16}<\partial^\mu \Phi,[A_\mu,\Phi]>,  \\
\\
\\
\frac{\textrm{g}}{4}(2\pi)^4C^{abc}\delta(p+q+r)(p_\mu-q_\mu)(g_{\alpha
\gamma}g_{\beta \delta}-g_{\beta \gamma}g_{\alpha \delta})
\end{array}
\qquad \qquad \qquad
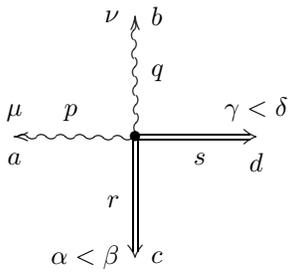
$$
 \vskip -2cm
$$
\xymatrix@!@=1.5cm{ & \ar@{<~}[d]^*+{q}_<*+{\nu}^<*+{b}& \\
 \ar@{<~}[r]^*+{p}^<*+{\mu}_<*+{a} &  *=0{\bullet}  & \ar@2{<-}[l]^*+{s}_<*+{\gamma < \delta}^<*+{d}\\
 & \ar@2{<-}[u]^*+{r}^<*+{\alpha < \beta}_<*+{c} & } \begin{array}{l} \\
\\
\\
\\
\\
 -\frac{i\textrm{g}^2}{32}<[A_\mu,\Phi],[A^\mu,\Phi]>,\\
\\
\\
-\frac{i\textrm{g}^2}{4}(2\pi)^4\delta(p+q+r+s)g_{\mu
\nu}(C^{eac}C^{ebd}+C^{ebc}C^{ead})\times \\
\\
\qquad \qquad \qquad \qquad \qquad \qquad \qquad \times(g_{\alpha
\gamma}g_{\beta \delta}-g_{\alpha \delta}g_{\beta \gamma})
\end{array}
$$
\vskip -2cm
\begin{equation}\label{cross2}
\hspace{-2.5cm}\xymatrix@!@=1.5cm{ & \ar@2{<-}[d]^*+{p}_<*+{\mu < \nu}^<*+{a}& \\
&  *=0{\bullet}  & \\
\ar@{<~}[ur]^*+{q}^<*+{\gamma}_<*+{b} & &
\ar@{<~}[ul]_*+{r}_<*+{\delta}^<*+{c}}
\hspace{-0.5cm}\begin{array}{l} \\
\\
\\
\\
\\
\frac{i\textrm{g}m}{8}<\Phi,[A\wedge,A]>, \\
\\
\\
\frac{i\textrm{g}m}{2}(2\pi)^4C^{abc}\delta(p+q+r)(g_{\mu
\gamma}g_{\nu \delta}-g_{\mu \delta}g_{\nu \gamma})
\end{array}
\end{equation}
\vskip -2cm
\begin{equation}\label{aee}
\hspace{-1.5cm}\xymatrix@!@=1.5cm{ & \ar@{<~}[d]_<*+{\mu}^<*+{a}^*+{k}& \\
&  *=0{\bullet}  & \\
\ar@{<--}[ur]_<*+{b}^*+{p} & & \ar@{>--}[ul]^<*+{c}_*+{q}}  \begin{array}{l} \\
\\
\\
\\
\\
-{i\textrm{g}}<d\overline \eta,[A,\eta]>,  \\
  \\
  \\
  -\textrm{g}C^{abc}(2\pi)^4\delta(k+p-q)p_\mu
\end{array}
\qquad \qquad \qquad \qquad
\end{equation}
\vskip -2cm
$$\hspace{-0.5cm}\xymatrix@!@=1.5cm{ & \ar@{<~}[d]_<*+{\mu}^<*+{a}^*+{k}& \\
&  *=0{\bullet}  & \\
\ar@{<--}[ur]_<*+{b}^*+{p}_{i} & & \ar@{>--}[ul]^<*+{c}_*+{q}^{i}}  \begin{array}{l} \\
\\
\\
\\
\\
-2{i\textrm{g}}<d\overline \eta_i,[A,\eta_i]>,  \\
  \\
  \\
  -2\textrm{g}C^{abc}(2\pi)^4\delta(k+p-q)p_\mu
\end{array}
\qquad \qquad \qquad \qquad \qquad
$$
\vskip -2cm
$$
\hspace{-0.5cm}\xymatrix@!@=1.5cm{ & \ar@{<~}[d]^*+{q}_<*+{\nu}^<*+{b}& \\
 \ar@{<~}[r]^*+{p}^<*+{\mu}_<*+{a} &  *=0{\bullet}  &
 \ar@{>--}[l]^*+{r}_{i}^<*+{c}\\
 & \ar@{<--}[u]^*+{s}_{i}_<*+{d} & } \begin{array}{l} \\
\\
\\
\\
\\
 {i\textrm{g}^2}<[A,\overline{\eta}_i]\wedge,[A,\eta_i]>,\\
\\
\\
2{i\textrm{g}^2}(2\pi)^4\delta(p+q+r+s)g_{\mu
\nu}(C^{eac}C^{ebd}+C^{ebc}C^{ead})
\end{array}
$$
In the coordinate formulas above $C^{abc}$ are the structure
constants of the Lie algebra $\mathfrak g$,
$[t^a,t^b]=C^{abc}t^c$. In the diagrams containing ghost lines we
assume that the variables $\eta$, $\eta_i$ correspond to the
outgoing ghost lines. All the coordinate formulas include the
corresponding combinatorial factors. Each diagram constructed with
the help of vertices and lines given above gives a contribution to
the generating function $Z(J)$. To calculate the contribution of a
connected diagram one has to integrate the expression
corresponding to the diagram with the measure
$\frac{d^4k}{(2\pi)^4}$ over all internal momenta and multiply the
result by $\frac{(-1)^l}{S}$, where $l$ in the number of the
$\eta$--\quad and $\eta_i$--ghost loops and $S$ is the order of
the symmetry group of the diagram.

\section{Effective action at the one-loop order and one-loop
renormalization}
In order to calculate the Green functions at the
one-loop order we shall use a formula expressing the effective
action at the one-loop order via a Gaussian Feynman path integral
(see \cite{IZ}, Sect. 12.3). We recall that the effective action
$\Gamma(A)$ is defined as the Legendre transform of the generating
function $G(J)=\ln Z(J)$,
$$
i\Gamma(A)=G(J)-i<J,A>,~A=\frac{\delta G}{i \delta J}.
$$
At the lowest order the effective action coincides with the
classical action of the theory. The first quantum one-loop
correction $\Gamma^1(A)$ can be represented as follows
\begin{eqnarray}\label{g1}
i\Gamma^1(A)= \qquad \qquad \qquad \qquad \qquad \qquad \qquad \qquad \qquad \qquad \qquad \qquad
\qquad \qquad \qquad \\
=\ln\left\{ \int{\mathcal D}(A'){\mathcal D}(\Phi'){\mathcal
D}(\eta'){\mathcal D}(\overline{\eta}')\prod_{i=1}^3{\mathcal
D}(\eta'_i){\mathcal D}(\overline{\eta}'_i) \exp
\{iS^2(A;A',\Phi',\eta',\overline{\eta}',\eta'_i,\overline{\eta}'_i)\}
\right\}, \nonumber
\end{eqnarray}
where
$S^2(A;A',\Phi',\eta',\overline{\eta}',\eta'_i,\overline{\eta}'_i)$
is the quadratic term in the Taylor expansion of the classical
action
$S(A+A',\Phi',\eta',\overline{\eta}',\eta'_i,\overline{\eta}'_i)$,
with respect to the primed variables at point $(A,0,0,0,0,0)$.

From formula (\ref{actloc}) using the definition of the curvature
we deduce that
\begin{eqnarray}
S^2= \frac{\textrm{g}}{2}<*[*F(A)\wedge,A'],A'>
-\frac{1}{2}<d_A*d_A A',A'>+ \qquad \nonumber \\
+\frac{1}{32}<\square_A\Phi',\Phi'> -\frac{m}{4}<\Phi',d_A
A'>+<\overline{\eta}',d^*d_A\eta'>- \label{s2}
\\-\sum_{i=1}^3<\overline{\eta}_i', \square_A\eta_i'>
-\frac{\lambda}{2}<dd^*A',A'>-\frac{m^2}{2}<d\square^{-1}d^*A',A'>.
\nonumber
\end{eqnarray}

Now Gaussian integrals in formula (\ref{g1}) can be easily
evaluated. This leads to the following expression for
$\Gamma^1(A)$
\begin{eqnarray}\label{g11}
  i\Gamma^1(A)=\ln \{ \det(-\square^{-1}d^*d_A)|_{\Omega^0(\mathfrak g)}\left( \det\left[
  (\square+m^2+(1-\lambda)dd^*)^{-1} \times \right. \right. \qquad \qquad \qquad \\
   \left. \left. \times(\textrm{g}*[*F(A)\wedge,\cdot~]
-d_A^*d_A-m^2d_A^*\square_A^{-1}d_A
-{\lambda}dd^*-{m^2}d\square^{-1}d^*)|_{\Omega^1(\mathfrak g)}
\right] \right)^{-\frac{1}{2}}\}. \nonumber
\end{eqnarray}
As usual we normalize the one-loop effective action in such a way
that $\Gamma^1(A)=0$ when $\textrm{g}=0$. Recalling that $\ln \det
X=\textrm{tr}\ln X$ for any operator $X$ we can rewrite formula
(\ref{g11}) in a form suitable for calculations,
\begin{eqnarray}\label{g12}
  i\Gamma^1(A)=\textrm{tr} \{ \ln(-\square^{-1}d^*d_A)|_{\Omega^0(\mathfrak g)} -\frac{1}{2}\ln\left[
  (\square+m^2+(1-\lambda)dd^*)^{-1} \times \right.  \qquad \qquad \qquad \\
  \left. \times(\textrm{g}*[*F(A)\wedge,\cdot~]
-d_A^*d_A-m^2d_A^*\square_A^{-1}d_A
-{\lambda}dd^*-{m^2}d\square^{-1}d^*)|_{\Omega^1(\mathfrak g)}
\right] \} . \nonumber
\end{eqnarray}
Formula (\ref{g12}) is the most explicit general expression for
the effective action at the one-loop order. The first term in the
r.h.s. of (\ref{g12}) is the contribution related to the
Faddeev-Popov determinant.

Now using formula (\ref{g12}) we shall calculate the one-loop
correction $\Gamma^1_2$ to the two-point function for the gauge
field $A$. In practical calculations we shall use the momentum
representation.

First we apply the Wick rotation to the time components of all
momenta and of the gauge field entering formula (\ref{g12}),
$$
k^0\mapsto ik^0,~A^0(k)\mapsto iA^0(k).
$$
We shall use dimensional regularization to preserve gauge
invariance, so that all the integrals with respect to momenta are
taken over ${\mathbb{R}}^d$ with respect to the measure
$\frac{d^dk}{(2\pi)^d}$. In the regularized expressions evaluated
in ${\mathbb{R}}^d$ the coupling constant $\textrm{g}$ should be
replaced with $\textrm{g}\mu^{\frac{\varepsilon}2},
\varepsilon=4-d$, where $\mu$ is an arbitrary mass scale, so that
$\textrm{g}$ remains dimensionless in ${\mathbb{R}}^d $.

The contribution $(\Gamma^1_2)_{{ghost}}$ to the two-point
function related to the Faddeev-Popov determinant
$$\textrm{tr} \{
\ln(-\square^{-1}d^*d_A)|_{\Omega^0(\mathfrak g)}\}=\textrm{tr} \{
\ln(I-\textrm{g}\square^{-1}\partial_\mu
[A^\mu,\cdot~])|_{\Omega^0(\mathfrak g)}\}
$$
in the r.h.s. of (\ref{g12}) is standard and coincides with that
for the pure Yang-Mills field. Indeed, expanding the r.h.s. of the
last formula with respect to $\textrm{g}$ we obtain
\begin{equation}\label{gh1}
i(\Gamma^1_2)_{{ghost}}=-\frac{\textrm{g}^2}{2}\textrm{tr}\left(\square^{-1}\partial_\mu\circ
[A^\mu,\cdot~]\circ\square^{-1}\partial_\nu \circ
[A^\nu,\cdot~]|_{\Omega^0(\mathfrak g)} \right),
\end{equation}
where the symbol $\circ$ stands for the composition of operators.
The trace in formula (\ref{gh1}) can be evaluated using the
momentum representation, with the conventions about the Wick
rotation and the dimensional regularization introduced above,
\begin{equation}\label{gh}
i(\Gamma^1_2)_{{ghost}}=\frac{\textrm{g}^2\mu^\varepsilon}{2}C\int\frac{(p+q)\cdot
\widehat A^a(p) ~q\cdot \widehat
A^a(-p)}{(p+q)^2q^2}\frac{d^dp}{(2\pi)^d}\frac{d^dq}{(2\pi)^d},
\end{equation}
where $\cdot$ denotes the standard scalar product of vectors in
$d$-dimensional Euclidean space ${\mathbb{R}}^d$, $q^2=q\cdot q$,
etc., and the constant $C$ is defined with the help of the
structure constants of the Lie algebra $\mathfrak g$,
$$
C\delta^{ab}=\sum_{c,d}C^{acd}C^{bcd}.
$$

The contribution $(\Gamma^1_2)_{A}$ to the two-point function
related to the second term
\begin{eqnarray}\label{t2}
\textrm{tr} \{ -\frac{1}{2}\ln[
  (\square+m^2+(1-\lambda)dd^*)^{-1}
  (\textrm{g}*[*F(A)\wedge,\cdot~]
-d_A^*d_A- \\
-m^2d_A^*\square_A^{-1}d_A
-{\lambda}dd^*-{m^2}d\square^{-1}d^*)|_{\Omega^1(\mathfrak g)} ]
\} \nonumber
\end{eqnarray}
in the r.h.s. of (\ref{g12}) can be calculated in a similar way.
For simplicity we shall only consider the case when $\lambda=1$.
Expanding the r.h.s. of (\ref{t2}) with respect to $\textrm{g}$
and using the resolvent formula for the d'Alambert operator
$\square_A^{-1}$,
$$
\square_A^{-1}=(I-\textrm{g}\square^{-1}(\partial_\mu\circ
[A^\mu,\cdot~]+ [A^\nu,\cdot~]\circ
\partial_\nu)+\textrm{g}^2\square^{-1}\circ[A^\mu,\cdot~]\circ
[A_\mu,\cdot~])^{-1}\square^{-1},
$$
we deduce that
\begin{equation}\label{ga}
i(\Gamma^1_2)_{A}=\frac{\textrm{g}^2}{2}\textrm{tr}(\frac{1}{2}X_1^2-X_2)|_{\Omega^1(\mathfrak
g)},
\end{equation}
where the operators $X_1$ and $X_2$ are defined by
\begin{eqnarray}\nonumber
X_1=(\square+m^2)^{-1}\left( d^*\circ
[A\wedge,\cdot~]+[A\wedge,\cdot~]^*\circ d- \right. \qquad \qquad
\qquad \qquad \qquad
\\
-m^2d^*\square^{-1}(\partial_\mu\circ [A^\mu,\cdot~]+
[A^\nu,\cdot~]\circ
\partial_\nu)\square^{-1}d+m^2[A\wedge,\cdot~]^*\circ
\square^{-1}d+ \label{x1} \\
\left. +m^2d^*\square^{-1}\circ
[A\wedge,\cdot~]+*[*(dA)\wedge,\cdot~] \right), \nonumber
\end{eqnarray}
\begin{eqnarray}\nonumber
X_2=(\square+m^2)^{-1}(-[A\wedge,\cdot~]^*\circ
[A\wedge,\cdot~]+m^2d^*\square^{-1}\circ[A^\mu,\cdot~]\circ
[A_\mu,\cdot~]\circ\square^{-1}d-
\\-m^2d^*\square^{-1}(\partial_\mu\circ [A^\mu,\cdot~]+
[A^\mu,\cdot~]\circ
\partial_\mu)\square^{-1}(\partial_\nu\circ [A^\nu,\cdot~]+
[A^\nu,\cdot~]\circ
\partial_\nu)\square^{-1}d+ \nonumber \\
+m^2[A\wedge,\cdot~]^*\circ \square^{-1}(\partial_\mu\circ
[A^\mu,\cdot~]+ [A^\nu,\cdot~]\circ
\partial_\nu)\square^{-1}d+ \label{x2} \\
+m^2d^*\square^{-1}(\partial_\mu\circ [A^\mu,\cdot~]+
[A^\nu,\cdot~]\circ
\partial_\nu)\square^{-1}\circ [A\wedge,\cdot~]-
\nonumber \\
-m^2[A\wedge,\cdot~]^*\circ \square^{-1}\circ
[A\wedge,\cdot~]-\frac{1}{2}*[*[A\wedge,A],\cdot~]). \nonumber
\end{eqnarray}
In formulas (\ref{x1}), (\ref{x2}) the superscript $*$ for
operators always denotes the operator adjoint with respect to the
scalar product on $\Omega^1(\mathfrak g)$.

Now we have to calculate the traces of the operators $X_1^2$ and
$X_2$ using the dimensional regularization as above. After tedious
algebra we get
\begin{eqnarray}
\textrm{tr}X_1^2=-C\int\frac{d^dp}{(2\pi)^d}\frac{d^dq}{(2\pi)^d}\{
2(d-1)\frac{(p+q)\cdot\widehat A^a(p)~q\cdot\widehat
A^a(-p)}{(p+q)^2q^2}+ \qquad \qquad \qquad \nonumber
\\
+m^4\frac{(p+2q)\cdot \widehat A^a(p)~(p+2q)\cdot \widehat
A^a(-p)((d-2)((p+q)\cdot
q)^2+(p+q)^2q^2)}{(p+q)^4q^4((p+q)^2+m^2)(q^2+m^2)}- \nonumber \\
-4m^2\frac{(p+2q)\cdot\widehat A^a(p)((d-2)q\cdot \widehat
A^a(-p)~(p+q)\cdot q-q^2(p+q)\cdot \widehat
A^a(-p))}{q^4(p+q)^2((p+q)^2+m^2)}+ \nonumber \\
+2\frac{(q^2+m^2)}{q^4((p+q)^2+m^2)}((d-2)q\cdot\widehat
A^a(-p)~q\cdot\widehat A^a(p)+q^2\widehat A^a(-p)\cdot\widehat
A^a(p))- \label{trx1} \\
-2m^2\frac{(p+2q)\cdot \widehat A^a(p)(q\cdot \widehat
A^a(-p)p^2-p\cdot q~p\cdot \widehat
A^a(-p))}{(p+q)^2q^2(q^2+m^2)((p+q)^2+m^2)} +\nonumber
\\
+4\frac{p\cdot \widehat A^a(p)~q\cdot \widehat A^a(-p)-p\cdot
q~\widehat A^a(p)\cdot \widehat A^a(-p)}{q^2((p+q)^2+m^2)}-
\nonumber
\\
-2\frac{p\cdot \widehat A^a(p)~p\cdot \widehat A^a(-p)-p^2\widehat
A^a(p)\cdot \widehat A^a(-p)}{((p+q)^2+m^2)(q^2+m^2)} \},\nonumber
\end{eqnarray}
\begin{eqnarray}
\textrm{tr}X_2=C(d-1)\int\frac{d^dp}{(2\pi)^d}\frac{d^dq}{(2\pi)^d}\{
\frac{m^2(p+q)^2-m^2q^2-q^2(p+q)^2}{(p+q)^2q^2(q^2+m^2)}\widehat
A^a(p)\cdot \widehat A^a(-p)- \label{trx2} \\
-m^2\frac{p\cdot \widehat A^a(-p)~(2q+p)\cdot \widehat
A^a(p)}{(p+q)^2q^2(q^2+m^2)}\}. \nonumber
\end{eqnarray}

As a function of $\varepsilon$, $\Gamma^1_2$ has a simple pole
singularity when $\varepsilon =4-d \rightarrow 0$. To find the
counterterm corresponding to the function $\Gamma^1_2$ we have to
calculate its residue at point $\varepsilon=0$ using formulas
(\ref{gh}), (\ref{ga}), (\ref{trx1}) and (\ref{trx2}).

Note that we use the minimal subtraction procedure to regularize
the effective action. Therefore the divergent part of
$\Gamma^1_2$, that we use to find the counterterms, is the
principal part of the Laurent series of $\Gamma^1_2(\varepsilon)$
at point $\varepsilon =0$.

The integrals over $q$ in (\ref{gh}), (\ref{trx1}) and
(\ref{trx2}) can be evaluated with the help of the following
formulas
\begin{eqnarray}\label{i1}
\int\frac{d^dq}{(2\pi)^d}\frac{1}{((p-q)^2+m_1^2)^n(q^2+m^2_2)^p}=\qquad
\qquad \qquad \qquad \qquad \qquad \qquad \qquad \qquad
\\
=\frac{1}{(4\pi)^{\frac{d}{2}}}\frac{\Gamma(n+p-\frac{d}{2})}{\Gamma(n)\Gamma(p)}
\int_{0}^{1}d\alpha
\alpha^{n-1}(1-\alpha)^{p-1}(\alpha(1-\alpha)p^2+\alpha m_1^2
+(1-\alpha)m_2^2)^{\frac{d}{2}-n-p}, \nonumber
\end{eqnarray}
\begin{eqnarray}\label{i2}
\int\frac{d^dq}{(2\pi)^d}\frac{q_\mu}{((p-q)^2+m_1^2)^n(q^2+m^2_2)^p}=\qquad
\qquad \qquad \qquad \qquad \qquad \qquad \qquad \qquad
\\
=\frac{p_\mu}{(4\pi)^{\frac{d}{2}}}\frac{\Gamma(n+p-\frac{d}{2})}{\Gamma(n)\Gamma(p)}
\int_{0}^{1}d\alpha
\alpha^n(1-\alpha)^{p-1}(\alpha(1-\alpha)p^2+\alpha m_1^2
+(1-\alpha)m_2^2)^{\frac{d}{2}-n-p}, \nonumber
\end{eqnarray}
\begin{eqnarray}\label{i3}
\int\frac{d^dq}{(2\pi)^d}\frac{q_\mu
q_\nu}{((p-q)^2+m_1^2)^n(q^2+m^2_2)^p}=\qquad \qquad \qquad \qquad
\qquad \qquad \qquad \qquad \qquad
\\*
=\frac{1}{(4\pi)^{\frac{d}{2}}}\{
\frac{\Gamma(n+p-\frac{d}{2})}{\Gamma(n)\Gamma(p)} \int_0^1d\alpha
\alpha^{n+1}(1-\alpha)^{p-1}(\alpha(1-\alpha)p^2+\alpha m_1^2 +
\nonumber\\* +(1-\alpha)m_2^2)^{\frac{d}{2}-n-p}p_\mu p_\nu+
\nonumber \\*
+\frac{\Gamma(n+p-\frac{d}{2}-1)}{\Gamma(n)\Gamma(p)}
\int_0^1d\alpha
\alpha^{n-1}(1-\alpha)^{p-1}(\alpha(1-\alpha)p^2+\alpha m_1^2 +
\nonumber
\\*
+(1-\alpha)m_2^2)^{\frac{d}{2}+1-n-p}\frac{\delta_{\mu \nu}}{2}
\}, \nonumber
\end{eqnarray}
where $\Gamma$ is the Euler gamma function.

Finally recalling that for $\varepsilon \rightarrow 0$ the
principal part of the Laurent series of the function $
\Gamma(-n+\varepsilon)$, $n=0,1,2, \ldots$ is equal to
$\frac{(-1)^n}{n!\varepsilon}$ we obtain an expression for the
divergent part $(\Gamma^1_2)_{div}$ of the function $\Gamma^1_2$,
\begin{eqnarray*}
i(\Gamma^1_2)_{div}=-\frac{C\textrm{g}^2}{(4\pi)^2\varepsilon}
\int\frac{d^dp}{(2\pi)^d}(\frac{5}{3}(p^2\widehat A^a(p)\cdot
\widehat A^a(-p)-p\cdot \widehat A^a(p)~p\cdot \widehat A^a(-p))+
\\* +2m^2\widehat A^a(p)\cdot \widehat A^a(-p)).
\end{eqnarray*}

Returning to the Minkowski space by the inverse Wick rotation we
can rewrite the last formula in the configuration space as follows
\begin{equation}\label{gdiv}
(\Gamma^1_2)_{div}=\frac{C\textrm{g}^2}{(4\pi)^2\varepsilon}
(\frac{5}{3}<dA,dA>+2m^2<A,A>).
\end{equation}

The divergent term (\ref{gdiv}) can be eliminated by adding a
counterterm $S^1_2$ to nonlocal action (\ref{anonloc}),
\begin{equation}\label{count}
S^1_2=-\frac{C\textrm{g}^2}{(4\pi)^2\varepsilon}
(\frac{5}{3}<dA,dA>+2m^2<A,A>).
\end{equation}
This is, of course, equivalent to a modification of the
coefficients of local action (\ref{actloc}). Actually only the
coefficients in the expression for the free action (\ref{actloco})
have to be renormalized, and the expression for the renormalized
free action $(S_0)^1_2$ takes the form
\begin{eqnarray}
(S_0)^1_2= -\frac{1}{2}Z_A<dA,dA>+\frac{1}{32}<\square\Phi,\Phi>
-\frac{m}{4}Z_m<\Phi,dA>+<d\overline{\eta},d\eta>+
\label{actloco1}
\\+\sum_{i=1}^3<d\overline{\eta}_i, d\eta_i>
-\frac{1}{2}<d^*A,d^*A>-\frac{m^2}{2}Z_m^2<\square^{-1}d^*A,d^*A>,
\nonumber
\end{eqnarray}
where
\begin{equation}\label{consts}
Z_A=1+\frac{\textrm{g}^2}{16\pi^2}\frac{5}{3}C\frac{2}{\varepsilon},
~Z_m=\sqrt{ 1-\frac{\textrm{g}^2}{16\pi^2}C\frac{4}{\varepsilon}}.
\end{equation}

By rescaling field $A$ in formula (\ref{actloco1}), $A^0=
Z_A^{\frac{1}{2}}A$, we obtain
\begin{eqnarray}
(S_0)^1_{2}= \label{actrenorm}
-\frac{1}{2}<dA^0,dA^0>+\frac{1}{32}<\square\Phi,\Phi>
-\frac{m}{4}Z_mZ_A^{-\frac{1}{2}}<\Phi,dA^0>+<d\overline{\eta},d\eta>+
\\+\sum_{i=1}^3<d\overline{\eta}_i, d\eta_i>
-\frac{Z_A^{-1}}{2}<d^*A^0,d^*A^0>-\frac{m^2}{2}Z_m^2Z_A^{-1}<\square^{-1}d^*A^0,d^*A^0>.
\nonumber
\end{eqnarray}

Now From formulas (\ref{consts}) and (\ref{actrenorm}) we can find
the mass renormalization,
\begin{equation}\label{mren}
m_0=mZ_mZ_A^{-\frac{1}{2}}=m\left(
1-\frac{\textrm{g}^2}{16\pi^2}\frac{11}{6}C\frac{2}{\varepsilon}\right),
\end{equation}
where $m_0$ is the bare mass.

Note that since the possible renormalization constant of the ghost
field $\Phi$ is canceled in the Gaussian integral in formula
(\ref{zo}) the renormalization of the wave function of $\Phi$ is
not required for calculating the mass renormalization.

To calculate the one-loop coupling constant renormalization one
has to find the divergent part of the three-point function
$\Gamma_3^1$ at the one-loop order. Looking at formulas (\ref{x1})
and (\ref{x2}) one should expect that technically this problem is
extremely complicated.

In the next section we shall prove the universality of the
coupling constant renormalization and Slavnov-Taylor identities
for the theory with generating function (\ref{gener}). Using the
universality of the coupling constant renormalization one can also
find the one-loop coupling constant renormalization with the help
of the one-loop corrections to the three-point function
represented by diagram (\ref{aee}). This calculation is standard
and is completely similar to the case of the pure Yang-Mills field
(see \cite{FS}, Ch. 4, \S 1). We only note that in case of the
dimensional regularization the mass in the propagator for the
gauge field does not add new divergent terms to the quantum
corrections. The coupling constant renormalization computed in
this way coincides with that for the pure Yang-Mills field,
\begin{equation}\label{gren}
\textrm{g}_0=\textrm{g}\left(
1-\frac{\textrm{g}^2}{16\pi^2}\frac{11}{6}C\frac{2}{\varepsilon}\right),
\end{equation}
where $\textrm{g}_0$ is the bare coupling constant.

One can also observe that the coupling constant renormalization in
the theory with generating function (\ref{gener}) calculated
within the dimentional regularization framework coincides with the
coupling constant renormalization for the pure Yang-Mills theory
to all orders of perturbation theory. Indeed, by the results of
\cite{th} in the case of dimensional regularization the coupling
constant renormalization is independent of the mass for
dimensional reasons. Therefore the coupling constant
renormalization in the theory with generating function
(\ref{gener}) is the same as in the massless case, i.e. in case of
the Yang-Mills field.

From formulas (\ref{consts}), (\ref{mren}) and (\ref{gren}) one
can calculate the corresponding renormalization group coefficients
at the one-loop order. An elementary renormalization group
analysis of the theory with generating function (\ref{gener}) can
be found in \cite{S}.

\section{Renormalization}

To complete our investigation of the theory with generating
function (\ref{gener}) we have to prove its renormalizability to
all orders of perturbation theory.

By simple dimensional counting all the integrals that enter the
perturbative expressions for the Green functions can be
dimensionally regularized. Indeed, the degree of divergence
$\omega(G)$ of any diagram $G$ that appears in the perturbative
expansion for generating function (\ref{gener}) can be computed as
follows:
\begin{equation}\label{deg}
\omega(G)=4-E-\delta+\sum(-1),
\end{equation}
where $E$ is the number of the external lines in the diagram $G$,
$\delta$ is the total power of external momenta factorized from
the corresponding integral, and the sum in the last term is taken
over parts (\ref{cross1}) of the free propagator which appear as
internal lines in $G$ and over vertexes (\ref{cross2}). From
formula (\ref{deg}) we deduce that divergences may only appear in
the Green functions with $E=2,3,4$, and the theory is
renormalizable.

We have to prove yet that the gauge invariance of the theory is
preserved by the renormalization. As usual this is achieved with
the help of the corresponding Slavnov-Taylor identities that we
are going to derive now.

The proof of renormalizability  of the theory with generating
function (\ref{gener}) is completely analogous to that for the
Yang-Mills field (see \cite{IZ}, Sect. 12.4). In the proof we
shall use the BRST technique. We shall consider the case of
arbitrary $\mathfrak g$-valued gauge fixing condition ${\mathcal
F}(A)$ linear in $A$, ${\mathcal F}(A)=\phi^\mu_{ab}A_\mu^b$. This
situation is slightly more general than the one discussed above
for action (\ref{actloc}), and the new total action including the
gauge fixing term takes the form
\begin{eqnarray}
S_{\mathcal F}= -\frac{1}{2}<F,F>+\frac{1}{32}<\square_A\Phi,\Phi>
-\frac{m}{4}<\Phi,F>-<\overline{\eta},\mathcal{M}\eta>+
\label{actlocf}
\\*+\sum_{i=1}^3<d_A\overline{\eta}_i, d_A\eta_i>
-\frac{\lambda}{2}<\mathcal{F}(A),\mathcal{F}(A)>, \nonumber
\end{eqnarray}
where $\mathcal M$ is the Faddeev-Popov operator for the gauge
fixing condition $\mathcal{F}(A)$.

The BRST transformation $s$ can be defined similarly to the case
of the Yang-Mills field (\cite{IZ}, Sect. 12.4.2),
\begin{equation}\label{brst}
\begin{array}{ll}
sA=d_A\eta, & s\overline \eta = \lambda \mathcal F(A),\\
   \\
s\Phi=\textrm{g}[\eta,\Phi], & s\eta_i=\textrm{g}[\eta,\eta_i],\\
  \\
s\eta=-\frac{\textrm{g}}{2}C^{abc}\eta^b\eta^c, & s\overline
\eta_i=\textrm{g}[\eta,\overline \eta_i].
\end{array}
\end{equation}
If we introduce the ghost number $\textrm{gh}$ for the fields as
follows
$$
\begin{array}{l}
\textrm{gh}A=\textrm{gh}\Phi=\textrm{gh}\eta_i=\textrm{gh}\overline \eta_i=0,  \\
\\
\textrm{gh}\eta =-1,~\textrm{gh}\overline \eta =1,
\end{array}
$$
then $s$ becomes a right superderivation of the algebra of the
fields, and
$$
s^2A=s^2\Phi=s^2\eta=s^2\eta_i=s^2\overline \eta_i=0, s^3\overline
\eta=0.
$$
We also define three ghost numbers $\textrm{gh}'_i$, $i=1,2,3$ by
$$
\begin{array}{l}
\textrm{gh}'_iA=\textrm{gh}'_i\Phi=\textrm{gh}'_i\eta=\textrm{gh}'_i\overline \eta=0,  \\
\\
\textrm{gh}'_i\eta_j =-\delta_{ij},~ \textrm{gh}'_i\overline
\eta_j =\delta_{ij}.
\end{array}
$$

We shall find the Slavnov-Taylor identities for generating
function $G(J,I,\xi,
\overline\xi,\xi^i,\overline{\xi}^i,K,L,N,P^i, \overline P^i)$,
\begin{eqnarray}
\exp G(J,I,\xi, \overline\xi,\xi^i,\overline{\xi}^i,K,L,N,P^i,
\overline P^i)=\int{\mathcal D}(A){\mathcal D}(\Phi){\mathcal
D}(\eta){\mathcal D}(\overline{\eta})\prod_{i=1}^3{\mathcal
D}(\eta_i){\mathcal D}(\overline{\eta}_i)\times \nonumber \\*
\times \exp \{i(S_{\mathcal
F}+<J,A>+\frac{1}{4}<I,\Phi>+<\overline\xi,\eta>+<\xi,
\overline\eta>+ \label{zf1} \\*
+\sum_{i=1}^3(<\overline{\xi}^i,\eta_i>+<\xi^i,
\overline{\eta}_i>)+<K,sA>-<L,s\eta>+\frac{1}{4}<N,s\Phi>+
\nonumber \\* +\sum_{i=1}^3(<P^i,s\eta_i>+<\overline
P^i,s\overline \eta_i>)\}, \nonumber
\end{eqnarray}
where $K,L,N,P^i, \overline P^i$ are the sources coupled to $sA$,
$s\eta$, $s\Phi$, $s\eta_i$ and $s\overline \eta_i$, respectively.
The generating function $G(J,I,\xi,
\overline\xi,\xi^i,\overline{\xi}^i,K,L,N,P^i, \overline P^i)$ is
reduced to $G(J)$ when all the sources, except for $J$, are equal
to zero. The new sources have the following ghost numbers
$$
\begin{array}{l}
\textrm{gh}K=\textrm{gh}N=\textrm{gh}P^i=\textrm{gh}\overline
P^i=1,~\textrm{gh}L=2,\\
\\
\textrm{gh}'_iK=\textrm{gh}'_iN=\textrm{gh}'_iL=0,~\textrm{gh}'_iP^j=\delta_{ij},
~\textrm{gh}'_i\overline P^j=-\delta_{ij}.
\end{array}
$$

We shall also need the dimensions of the fields and of the
sources,
$$
\begin{array}{l}
 \textrm{dim}A=\textrm{dim}\Phi=\textrm{dim}\eta=\textrm{dim}\overline \eta=
 \textrm{dim}\eta_i=\textrm{dim}\overline \eta_i=1, \\
 \\
 \textrm{dim}K=\textrm{dim}L=\textrm{dim}N=\textrm{dim}P_i=\textrm{dim}\overline
 P_i=2, \\
 \\
\textrm{dim}J=\textrm{dim}I=\textrm{dim}\xi=\textrm{dim}\overline
\xi=\textrm{dim}\xi^i=\textrm{dim}\overline \xi^i=3,
\end{array}
$$
the dimension of the mass $m$ being equal to 1, and the coupling
constant $\textrm{g}$ is dimensionless.

To derive the Slavnov-Taylor identities we have to apply the BRST
transformation (\ref{brst}) to the variables of integration in the
r.h.s. of formula (\ref{zf1}). Observe that both action
(\ref{actlocf}) and the measure in the Feynman path integral
(\ref{actlocf}) are BRST-invariant. The former statement can be
checked directly, and the latter one is true since the BRST
transformation is nilpotent. Therefore the  BRST coordinate change
in (\ref{actlocf}) yields the following identity
\begin{eqnarray*}
\int{\mathcal D}(A){\mathcal D}(\Phi){\mathcal D}(\eta){\mathcal
D}(\overline{\eta})\prod_{i=1}^3{\mathcal D}(\eta_i){\mathcal
D}(\overline{\eta}_i)(<J,sA>+\frac{1}{4}<I,s\Phi>+<\overline\xi,s\eta>- \quad \nonumber \\
-< s\overline\eta,\xi>+\sum_{i=1}^3(<\overline{\xi}^i,s\eta_i>+<
s\overline{\eta}_i,\xi^i>))\exp \{i(S_{\mathcal
F}+<J,A>+\frac{1}{4}<I,\Phi>+ \nonumber \\
+<\overline\xi,\eta>+<\xi,
\overline\eta>+\sum_{i=1}^3(<\overline{\xi}^i,\eta_i>+<\xi^i,
\overline{\eta}_i>)+<K,sA>-<L,s\eta>+  \\
+\frac{1}{4}<N,s\Phi>+\sum_{i=1}^3(<P^i,s\eta_i>+<\overline
P^i,s\overline \eta_i>)\}=0. \nonumber
\end{eqnarray*}
This identity can be rewritten as
\begin{eqnarray*}
(<J,\frac{\delta}{i\delta K}>+\frac{1}{4}<I,\frac{\delta}{i\delta
N}> -<\overline\xi,\frac{\delta}{i\delta L}>-\lambda<\xi,{\mathcal
F}(\frac{\delta}{i\delta
J})>+ \qquad \qquad \qquad \nonumber \\
+\sum_{i=1}^3(<\overline{\xi}^i,\frac{\delta}{i\delta P^i}>-<
\xi^i,\frac{\delta}{i\delta \overline P^i}>))\exp G(J,I,\xi,
\overline\xi,\xi^i,\overline{\xi}^i,K,L,N,P^i, \overline P^i) = 0
,
\end{eqnarray*}
which is, in turn, equivalent to
\begin{eqnarray}
(<J,\frac{\delta}{i\delta K}>+ \frac{1}{4}<I,\frac{\delta}{i\delta
N}>-<\overline\xi,\frac{\delta}{i\delta L}>-\lambda<\xi,{\mathcal
F}(\frac{\delta}{i\delta J})>+\qquad \qquad \qquad  \label{sl1}
\\*
 +\sum_{i=1}^3(<\overline{\xi}^i,\frac{\delta}{i\delta P^i}>-<
\xi^i,\frac{\delta}{i\delta \overline P^i}>))G(J,I,\xi,
\overline\xi,\xi^i,\overline{\xi}^i,K,L,N,P^i, \overline P^i)=0
\nonumber
\end{eqnarray}
due to linearity of $\mathcal F$.

The Slavnov-Taylor identities (\ref{sl1}) are supplemented with
another identity following from the fact that the Feynman path
integral in the r.h.s. of formula (\ref{zf1}) is invariant under
translations of the ghost variable of integration,
$$
\overline \eta \mapsto \overline \eta +\delta \overline \eta.
$$
From this fact we infer that
\begin{eqnarray*}
\int{\mathcal D}(A){\mathcal D}(\Phi){\mathcal D}(\eta){\mathcal
D}(\overline{\eta})\prod_{i=1}^3{\mathcal D}(\eta_i){\mathcal
D}(\overline{\eta}_i)(-{\mathcal M}\eta + \xi)\exp \{i(S_{\mathcal
F}+<J,A>+\frac{1}{4}<I,\Phi>+ \nonumber \\*
+<\overline\xi,\eta>+<\xi,
\overline\eta>+\sum_{i=1}^3(<\overline{\xi}^i,\eta_i>+<\xi^i,
\overline{\eta}_i>)+<K,sA>-<L,s\eta>+  \\*
+\frac{1}{4}<N,s\Phi>+\sum_{i=1}^3(<P^i,s\eta_i>+<\overline
P^i,s\overline \eta_i>)\}=0, \nonumber
\end{eqnarray*}
or, since ${\mathcal M}\eta=s{\mathcal F}=\phi^\mu_{ab}
(sA)_\mu^{b}:=\phi\cdot sA$,
$$
(\xi-\phi\cdot \frac{\delta}{i\delta K})\exp G(J,I,\xi,
\overline\xi,\xi^i,\overline{\xi}^i,K,L,N,P^i, \overline P^i)=0.
$$
For connected functions the last identity gives
\begin{equation}\label{sl2}
\phi\cdot \frac{\delta}{i\delta K}G(J,I,\xi,
\overline\xi,\xi^i,\overline{\xi}^i,K,L,N,P^i, \overline P^i)=\xi.
\end{equation}

In the proof of renormalizability it is more convinient to use the
proper functions instead of the Green functions, and instead of
the generating function $G(J,I,\xi,
\overline\xi,\xi^i,\overline{\xi}^i,K,L,N,P^i, \overline P^i)$ we
shall consider its Legendre transform $\Gamma(A,\Phi,\eta,
\overline\eta,\eta_i,\overline{\eta}_i,K,L,N,P^i, \overline P^i)$,
\begin{eqnarray} \Gamma(A,\Phi,\eta,
\overline\eta,\eta_i,\overline{\eta}_i,K,L,N,P^i, \overline
P^i)=-iG(J,I,\xi, \overline\xi,\xi^i,\overline{\xi}^i,K,L,N,P^i,
\overline P^i)- \quad \label{lege}
\\
-<J,A>+\frac{1}{4}<I,\Phi>+<\overline\xi,\eta>+<\xi,
\overline\eta>+\sum_{i=1}^3(<\overline{\xi}^i,\eta_i>+<\xi^i,
\overline{\eta}_i>), \nonumber
\end{eqnarray}
where
\begin{equation}\label{lege1}
A=\frac{\delta G}{i\delta J},~\eta=\frac{\delta G}{i\delta
\overline \xi},~\overline \eta=-\frac{\delta G}{i\delta
\xi},~\Phi=\frac{\delta G}{i\delta I},~\eta_i=\frac{\delta
G}{i\delta \overline \xi^i},~\overline \eta_i=-\frac{\delta
G}{i\delta \xi^i},
\end{equation}
and the variables $K,~L,~N,~P^i,~ \overline P^i$ are passive.

Since
$$
J=-\frac{\delta \Gamma}{\delta A},~\overline \xi=\frac{\delta
\Gamma}{\delta \overline \eta},~\xi=-\frac{\delta \Gamma}{\delta
\overline \eta},~I=-\frac{\delta \Gamma}{\delta \Phi},~\overline
\xi^i=\frac{\delta \Gamma}{\delta \overline
\eta_i},~\xi^i=-\frac{\delta \Gamma}{\delta \overline \eta_i}
$$
and
$$
\frac{\delta G}{i\delta K}=\frac{\delta \Gamma}{\delta
K},~\frac{\delta G}{i\delta L}=\frac{\delta \Gamma}{\delta
L},~\frac{\delta G}{i\delta N}=\frac{\delta \Gamma}{\delta
N},~\frac{\delta G}{i\delta P^i}=\frac{\delta \Gamma}{\delta
P^i},~\frac{\delta G}{i\delta \overline P^i}=\frac{\delta
\Gamma}{\delta \overline P^i}
$$
identities (\ref{sl1}), (\ref{sl2}) can be expressed in terms of
the function $\Gamma$ as
\begin{eqnarray}
<\frac{\delta \Gamma}{\delta A},\frac{\delta \Gamma}{\delta K}>+
\frac{1}{4}<\frac{\delta \Gamma}{\delta \Phi},\frac{\delta
\Gamma}{\delta N}>+<\frac{\delta \Gamma}{\delta \eta},\frac{\delta
\Gamma}{\delta L}>-\lambda<\frac{\delta \Gamma}{\delta \overline
\eta},{\mathcal F}(A)>+\qquad   \label{sl1g}  \\
+\sum_{i=1}^3(-<\frac{\delta \Gamma}{\delta \eta_i},\frac{\delta
\Gamma}{\delta P^i}>+<\frac{\delta \Gamma}{\delta \overline
\eta_i},\frac{\delta \Gamma}{\delta \overline P^i}>)=0, \nonumber
\end{eqnarray}
\begin{equation}\label{sl2g}
\phi\cdot \frac{\delta \Gamma}{\delta K}+\frac{\delta
\Gamma}{\delta \overline \eta}=0.
\end{equation}

Formulas (\ref{sl1g}), (\ref{sl2g}) take a slightly simpler form
for the modified effective action $\widetilde{\Gamma}$ defined by
$$
\widetilde{\Gamma}=\Gamma+\frac{\lambda}{2}<{\mathcal F},{\mathcal
F}>.
$$
Indeed, from (\ref{sl1g}), (\ref{sl2g}) we have
\begin{eqnarray}
<\frac{\delta \widetilde\Gamma}{\delta A},\frac{\delta
\widetilde\Gamma}{\delta K}>+ \frac{1}{4}<\frac{\delta
\widetilde\Gamma}{\delta \Phi},\frac{\delta
\widetilde\Gamma}{\delta N}>+<\frac{\delta
\widetilde\Gamma}{\delta \eta},\frac{\delta
\widetilde\Gamma}{\delta L}>+ \qquad \label{sl1g1} \\
+\sum_{i=1}^3(-<\frac{\delta \widetilde\Gamma}{\delta
\eta_i},\frac{\delta \widetilde\Gamma}{\delta P^i}>+<\frac{\delta
\widetilde\Gamma}{\delta \overline \eta_i},\frac{\delta
\widetilde\Gamma}{\delta \overline P^i}>)=0, \nonumber
\end{eqnarray}
$$
\phi\cdot \frac{\delta \widetilde\Gamma}{\delta K}+\frac{\delta
\widetilde \Gamma}{\delta \overline \eta}=0.
$$

For the operation in the r.h.s. of formula (\ref{sl1g1}) we shall
use the compact notation
\begin{eqnarray}
\Gamma_1*\Gamma_2=<\frac{\delta \Gamma_1}{\delta A},\frac{\delta
\Gamma_2}{\delta K}>+ \frac{1}{4}<\frac{\delta \Gamma_1}{\delta
\Phi},\frac{\delta \Gamma_2}{\delta N}>+<\frac{\delta
\Gamma_1}{\delta \eta},\frac{\delta
\Gamma_2}{\delta L}>+\qquad   \label{oper}  \\
+\sum_{i=1}^3(-<\frac{\delta \Gamma_1}{\delta \eta_i},\frac{\delta
\Gamma_2}{\delta P^i}>+<\frac{\delta \Gamma_1}{\delta \overline
\eta_i},\frac{\delta \Gamma_2}{\delta \overline P^i}>). \nonumber
\end{eqnarray}

Now we can prove that the theory with generating function
(\ref{gener}) is renormalizable. First recall that by simple
dimensional counting the integrals contained in the perturbative
loop expansion
\begin{equation}\label{exp}
\widetilde\Gamma=\widetilde\Gamma^0+\widetilde\Gamma^1+\widetilde\Gamma^2+\cdots
\end{equation}
for the generating function $\widetilde \Gamma$ can be
dimensionally regularized. Assume for a moment that this is done,
i.e. the Wick rotation is performed and all the integrals are
taken over ${\mathbb{R}}^d$ as in the one-loop case discussed in
the previous section. When the dimension $d$ of the space
${\mathbb{R}}^d$ goes to $4$ each term in expansion (\ref{exp})
acquires a divergent term corresponding to a pole in variable
$\varepsilon=4-d$ as $\varepsilon \rightarrow 0$ (minimal
subtraction). The divergent terms $\widetilde\Gamma^n_{div}$ can
be determined recursively from the identity
$$
\widetilde\Gamma^n_{reg}=\widetilde\Gamma^n_R+\widetilde\Gamma^n_{div},
$$
where $\widetilde\Gamma^n_{reg}$ is computed taking into account
all lower-order counterterms in the perturbation expansion, and
$\widetilde\Gamma^n_R$ is the renormalized effective action at the
n-th order.

At the zeroth order we obviously have
\begin{eqnarray}
\widetilde\Gamma^0=-\frac{1}{2}<F,F>+\frac{1}{32}<\square_A\Phi,\Phi>
-\frac{m}{4}<\Phi,F>-<\overline{\eta},\mathcal{M}\eta>+\nonumber
\\
+\sum_{i=1}^3<d_A\overline{\eta}_i, d_A\eta_i>
+<K,sA>-<L,s\eta>+\frac{1}{4}<N,s\Phi>+ \label{g0}
 \\
+\sum_{i=1}^3(<P^i,s\eta_i>+<\overline P^i,s\overline \eta_i>)=
\nonumber
\end{eqnarray}
\begin{eqnarray}
=-\frac{1}{2}<F,F>+\frac{1}{32}<\square_A\Phi,\Phi>
-\frac{m}{4}<\Phi,F>+<(K-\overline{\eta}\cdot \phi),sA>+\nonumber
\\
+\sum_{i=1}^3<d_A\overline{\eta}_i, d_A\eta_i>
-<L,s\eta>+\frac{1}{4}<N,s\Phi>+ \nonumber
 \\
+\sum_{i=1}^3(<P^i,s\eta_i>+<\overline P^i,s\overline \eta_i>),
\nonumber
\end{eqnarray}
where $\overline{\eta}\cdot \phi^\mu=\overline{\eta}^a
\phi_{ab}^\mu$.

From (\ref{sl1g1}) we infer by induction over $n$ that every
divergent part $\widetilde\Gamma^n_{div}$ must satisfy the
following equation (see \cite{IZ}, Sect. 12.4.3 for details):
\begin{equation}\label{diveq}
\widetilde\Gamma^0*\widetilde\Gamma^n_{div}+\widetilde\Gamma^n_{div}*\widetilde\Gamma^0=0.
\end{equation}
Equation (\ref{diveq}) allows to find the most general form of the
divergent terms.

If we assume, by the choice of gauge, that the symmetry of
$\widetilde\Gamma^n_{div}$ under constant gauge transformations is
unbroken then simple power counting and the ghost number
conservation restrict the form of $\widetilde\Gamma^n_{div}$ to
\begin{eqnarray}
\widetilde\Gamma^n_{div}={\mathcal
L}^n+\alpha<(K-\overline{\eta}\cdot
\phi),sA>+(\alpha-\beta)<(K-\overline{\eta}\cdot
\phi),[A,\eta]>+\label{gdivn}
\\
+\gamma<L,s\eta>+\delta \frac{1}{4}<N,s\Phi>+ \nonumber
+\sum_{i=1}^3(\varepsilon_i<P^i,s\eta_i>+\overline
\varepsilon_i<\overline P^i,s\overline \eta_i>),
\end{eqnarray}
where $\alpha$, $\beta$, $\gamma$, $\delta$, $\varepsilon_i$,
$\overline \varepsilon_i$ are numerical coefficients, and
${\mathcal L}^n$ is a functional of degree four depending on $A,
\Phi, \eta_i, \overline \eta_i$ and having zero ghost numbers.

Substituting expression (\ref{gdivn}) into (\ref{diveq}) we get
\begin{equation}\label{constseq}
\beta=\delta=\varepsilon_i=\overline \varepsilon_i=-\gamma,
\end{equation}
\begin{eqnarray}\label{leq}
(\alpha-\beta)<\frac{\delta {\mathcal L}}{\delta
A},d\eta>+<\frac{\delta {\mathcal L}^n}{\delta A},sA>
+\frac{1}{4}<\frac{\delta {\mathcal L}^n}{\delta \Phi}, s\Phi>-
\\
-<\frac{\delta {\mathcal L}^n}{\delta
\eta_i},s\eta_i>-<\frac{\delta {\mathcal L}^n}{\delta \overline
\eta_i},s\overline \eta_i>=0, \nonumber
\end{eqnarray}
where
$$
{\mathcal L}=-\frac{1}{2}<F,F>+\frac{1}{32}<\square_A\Phi,\Phi>
-\frac{m}{4}<\Phi,F>+\sum_{i=1}^3<d_A\overline{\eta}_i,
d_A\eta_i>.
$$

The general solution to equation (\ref{leq}) is
\begin{equation}\label{ln}
{\mathcal L}^n=(\beta-\alpha)(<\frac{\delta {\mathcal L}}{\delta
A},A>+\frac{1}{4}<\frac{\delta {\mathcal L}}{\delta \Phi}, \Phi>
-<\frac{\delta {\mathcal L}}{\delta \eta_i},\eta_i>-<\frac{\delta
{\mathcal L}}{\delta \overline \eta_i},\overline
\eta_i>)+{\mathcal L}_{inv},
\end{equation}
where ${\mathcal L}_{inv}$ is an arbitrary gauge invariant
functional of degree four depending on $A, \Phi, \eta_i, \overline
\eta_i$ and having zero ghost numbers.

We actually do not need to consider the most general expression
for ${\mathcal L}_{inv}$. Some terms in this expression are a
priori not required for the renormalization. From formula
(\ref{actloc}) it follows that for $m=0$  the spatial components
of the ghost tensor field $\Phi$ do not interact with each other
and with the other ghost fields. Taking into account the ghost
numbers conservation we infer that the counterterms containing
interactions of different components of $\Phi$ and of
anticommuting ghosts $\eta_i$, $\overline \eta_i$ for different
indexes i should appear with multiples $m$ or $\frac{m}{\mu}$. The
latter possibility is not realized in case of the minimal
subtraction. Note also that the ghosts $\eta_i$, $\overline
\eta_i$ symmetrically appear in formula (\ref{actloc}). Therefore
they must symmetrically appear in the expression for the
counterterms.

These rules restrict the expression for ${\mathcal L}_{inv}$
required for the renormalization to
\begin{eqnarray}
{\mathcal
L}_{inv}=-\frac{a_0}{2}<F,F>+\frac{a_1}{32}<\square_A\Phi,\Phi>
-\frac{a_3m}{4}<\Phi,F>+ \qquad \qquad \label{linv} \\*
+\sum_{i=1}^3a_2<d_A\overline{\eta}_i, d_A\eta_i>
+\sum_{i=1}^3a_4m_\eta^2<\overline
\eta_i,\eta_i>+a_5\frac{m_\Phi^2}{4}<\Phi,\Phi>, \nonumber
\end{eqnarray}
where $a_0,a_1,a_2$, $a_3$, $a_4,a_5$ are nimerical coefficients,
and four new mass parameters $m_\Phi$, $m_\eta$ are, of course,
proportional to $m$.

Substituting (\ref{ln}) and (\ref{linv}) into (\ref{gdivn}), using
relations (\ref{constseq}) and the homogeneity property of the
functional ${\mathcal L}_{inv}$ with respect to coupling constants
and fields we finally obtain the following expression for the most
general form of the counterterms (compare with \cite{IZ}, Sect
12.4.3., formula 12-165)
\begin{eqnarray}
\widetilde\Gamma^n_{div}=\{
(\beta-\alpha+\frac{a_0}{2})<A,\frac{\delta}{\delta
A}>+\frac{1}{4}(\beta-\alpha+\frac{a_1}{2})<\Phi,\frac{\delta}{\delta
\Phi}>+(a_3-\frac{a_0+a_1}{2})m\frac{\partial}{\partial m}+
 \nonumber
\\
+\frac{a_0-a_1}{8}<N,\frac{\delta}{\delta
N}>+\alpha<\eta,\frac{\delta}{\delta
\eta}>+\sum_{i=1}^3(\frac{a_0-a_2}{2})(<P_i,\frac{\delta}{\delta
P_i}>+<\overline P_i,\frac{\delta}{\delta \overline P_i}>)+ \label{fingd} \\
+\sum_{i=1}^3(\beta-\alpha+\frac{a_2}{2})(<\eta_i,\frac{\delta}{\delta
\eta_i}>+<\overline \eta_i,\frac{\delta}{\delta \overline
\eta_i}>)
+\sum_{i=1}^3(a_4-a_2-2(\beta-\alpha))m_\eta^2\frac{\partial}{\partial
m_\eta^2}+\nonumber \\
+(a_5-a_1-2(\beta-\alpha))m_\Phi^2\frac{\partial}{\partial
m_\Phi^2}-a_0g\frac{\partial}{\partial g}
+(\beta-2\alpha+\frac{a_0}{2})<L,\frac{\delta}{\delta L}>\nonumber
\}(\widetilde\Gamma^0)',\nonumber
\end{eqnarray}
where
\begin{eqnarray}
(\widetilde\Gamma^0)'=-\frac{1}{2}<F,F>+\frac{1}{32}<\square_A\Phi,\Phi>
-\frac{m}{4}<\Phi,F>+\sum_{i=1}^3<d_A\overline{\eta}_i,
d_A\eta_i>-\nonumber
\\
-<\overline{\eta},{\mathcal M}\eta>+\sum_{i=1}^3m_\eta^2<\overline
\eta_i,\eta_i>+ \label{gmod} \frac{m_\Phi^2}{4}<\Phi,\Phi>
+<K,sA>-<L,s\eta>+
 \\
+\frac{1}{4}<N,s\Phi>+\sum_{i=1}^3(<P^i,s\eta_i>+<\overline
P^i,s\overline \eta_i>). \nonumber
\end{eqnarray}

Formula (\ref{fingd}) tells that the divergent part of the
effective action at the n-th order has the same form as action
(\ref{gmod}) with renormalized coupling constants, masses and
fields. Comparing formula (\ref{gmod}) with (\ref{g0}) we see that
a priori new counterterms have to be added to the original action
(\ref{g0}) to cancel all the divergences. The only difference
between expressions (\ref{gmod}) and (\ref{g0}) is that the ghost
fields acquire masses.

The renormalized generating function $G_R(J)$ is obtained by
taking the inverse Legendre transform of the renormalized
effective action $\Gamma_R(A,\Phi,\eta,
\overline\eta,\eta_i,\overline{\eta}_i,K,L,N,P^i, \overline P^i)$
and by putting all the arguements, except for $J$, equal to zero.
Recalling the definition (\ref{lege}), (\ref{lege1}) of the
Legendre transform we get
\begin{eqnarray*}
G_R(J,I,\xi, \overline\xi,\xi^i,\overline{\xi}^i,K,L,N,P^i,
\overline P^i)=i\Gamma_R(A,\Phi,\eta,
\overline\eta,\eta_i,\overline{\eta}_i,K,L,N,P^i, \overline P^i)+
\quad
\\
+i(<J,A>-\frac{1}{4}<I,\Phi>-<\overline\xi,\eta>-<\xi,
\overline\eta>-\sum_{i=1}^3(<\overline{\xi}^i,\eta_i>+<\xi^i,
\overline{\eta}_i>)),
\end{eqnarray*}
$$
G_R(J)=G_R(J,0,0, 0,0,0,0,0,0,0, 0).
$$
This completes the perturbative study of the theory with
generating function (\ref{gener}).

\vskip 0.5cm

{\large \bf Acknowledgements.} When this work was completed we
have received information that the action similar to (\ref{act1})
was also recently introduced and studied in \cite{C}. Paper
\cite{C} also contains an extensive list of references on
dynamical mass generation in the Yang-Mills theory.

Prof. R. Jackiw also communicated to the author of this paper that
in case of three dimensions the mass generation mechanism similar
to the one introduced in this paper was suggested in \cite{Jw}.
The author would like to thank him and  David Dudal for pointing
out the references quoted above.

The author is also greatly indebted to John A. Gracey for an
illuminating discussion of the mass renormalization.


\begin{thebibliography}{99}

\bibitem{C} Capri, M. A. L., Dudal, D., Gracey, J. A., Lemes, V. E. R., Sobreiro, R. F.,
Sorella, S. P., Verschelde, H.: Study of the gauge invariant,
nonlocal mass operator $\textrm{Tr}\int
d^4xF_{\mu\nu}(D^2)^{-1}F_{\mu\nu}$ in Yang-Mills theories, {\em
Phys. Rev.} {\bf D 72} (2005), 105016.

\bibitem{FP} Faddeev, L. D., Popov, V. N.: Feynman
diagrams for the Yang-Mills field, {\em Phys. Lett.}  {\bf B 25}
(1967), 29.

\bibitem{FS} Faddeev, L. D., Slavnov, A. A.: Gauge fields.
Introduction to quantum theory, {\em Frontiers in Physics, 83},
Addison-Wesley (1991).

\bibitem{th} Hooft, G. 't: Dimensional regularization and the
renormalization group, {\em Nucl. Phys.} {\bf B 61} (1973),
455--468.

\bibitem{IZ} Itzykson, C., Zuber, J.-B.: Quantum Field Theory,
McGraw-Hill (1980).

\bibitem{Jw} Jackiw, R., Pi, S.-Y.: Seeking an even-parity mass
term for 3-D gauge theory, {\em Phys. Lett.} {\bf B 403} (1997),
297.

\bibitem{R} Ramond, P.: Field theory: a modern primer,
Addison-Wesley (1989).

\bibitem{S} Sevostyanov, A.: A mass generation mechanism for gauge
fields, preprint hep-th/0605050.

\end{thebibliography}
\end{document}